\documentclass[aps,PRB,reprint,superscriptaddress,preprintnumbers]{revtex4-2}
\usepackage[T1]{fontenc}
\usepackage{times}
\usepackage{graphicx,color}
\usepackage{amsfonts,amsmath,amssymb,amsbsy}
\usepackage[colorlinks=true, linkcolor=blue, citecolor=blue, urlcolor=blue]{hyperref}
\usepackage{float}
\usepackage{lettrine}
\let\bm=\boldsymbol
\begin{document}

\title{Nodal-line semimetals and their variance}

\author{Po-Yao Chang}
\email{pychang@phys.nthu.edu.tw} 
\affiliation{Department of Physics, National Tsing Hua University, Hsinchu 30013, Taiwan}
\affiliation{Yukawa Institute for Theoretical Physics, Kyoto University, Kyoto 606-8502, Japan}

\clearpage

\begin{abstract}
Topological nodal-line semimetals (NLSMs) are a new family of topological materials characterized by electronic band crossings that form lines in the Brillouin zone.
These NLSMs host exotic nodal-line structures and exhibit distinct features such as drumhead surface states and unique electromagnetic responses. 
This review classifies various NLSM types based on their nodal structures and protecting symmetries, highlighting that these nodal-line structures can form links, knots, and chains. 
We discuss their characteristic electromagnetic responses, including Landau level spectroscopy, optical conductivity, and permittivity. Furthermore, the strong correlation effects in these NLSMs modify their semimetallic phases and lead to novel quantum phases where magnetism and superconductivity intertwine.

\end{abstract}

\date{\today}

\maketitle

\clearpage
%\tableofcontents

\section{Introduction}
%The discovery of topological materials reveal exotic phases of matter emerging from intertwine with quantum phenomena and topological properties.
%The dispersionless boundary states of topological insulators are immune under disorders~\cite{Hasan2010, Qi2011, YAN2025100023}. The unusual electromagnetic responses in the Weyl semimetals can be observed from
%the reduction of the magnetoresistance due to the chiral magnetic effect. These topological properties are potentially applicable for next generation devices which drew a huge attention across all fields.

The quest for novel quantum states of matter, driven by the intricate interplay of symmetry, topology, and quantum mechanics, is a central theme in modern condensed matter physics. Among the most fascinating discoveries in recent decades is the emergence of topological materials, which host exotic phases characterized by unique electronic structures and protected boundary phenomena. Prominent examples include topological insulators (TIs), featuring insulating bulk states but conductive surface states whose transport properties are remarkably robust against non-magnetic disorder, offering potential pathways towards dissipationless electronics~\cite{Hasan2010, Qi2011, YAN2025100023}. Another key family is Weyl semimetals (WSMs), characterized by bulk band-touching points (Weyl nodes) that act as sources or sinks of Berry curvature, leading to unique electromagnetic phenomena rooted in fundamental chiral anomalies, observable through signatures like the chiral magnetic effect-induced negative magnetoresistance~\cite{Armitage2018}. The profound physics and potential application of these topological properties in areas ranging from low-power electronics to quantum computing have spurred significant theoretical and experimental interest across diverse research fields.

%A new family of topological materials was first introduced in Ref.~\cite{Burkov2011}. The single-particle energy spectra of these materials have gap closing points forming lines in the three dimensional Brillouin zone (BZ).
%These materials are referred to as the topological nodal-line semimetals (NLSMs).
%NLSMs are similar as their topological family members, existing protected surface states and unusual electromagnetic responses are also exhibit in these materials.
%The protected surface states are originated from the bulk nodal lines projected on the surface BZ~\cite{Burkov2011, Matsuura_2013, Yu2015, Bian2016,Chan2016}.
%There are also zeroth Landau level physics in the NLSMs which can be heuristically understood from the layered graphene structures~\cite{Rhim2015, Chang2017}.
%Experimentally, the drumhead states have been observed from ARPES~\cite{Bian2016}.

Beyond the point-like nodes of WSMs, a distinct class of topological gapless systems emerged: topological nodal-line semimetals (NLSMs), first proposed theoretically in Ref.~\cite{Burkov2011}. These materials are distinguished by their electronic band structures where conduction and valence bands touch not at isolated points, but along continuous lines or loops within the three-dimensional Brillouin zone (BZ). This distinct nodal structure, protected by symmetries like mirror reflection or time-reversal combined with inversion, imparts unique characteristics. Similar to their topological counterparts, NLSMs possess protected surface states, which often manifest as two-dimensional 'drumhead' states spanning the region enclosed by the projection of the bulk nodal lines onto the surface BZ~\cite{Burkov2011, Matsuura_2013, Yu2015, Bian2016, Chan2016, Zhao2013, Fang2015, Fang_2016, Zhao2017, Lim2017, Sun_2017, Wang2017, Ratnadwip2017, Pan2018,Guo2019, Hirose2020,Okamoto2020, Stuart2022, Gao2023, Jeon2023}. These surface states represent a key experimental fingerprint. Furthermore, NLSMs can exhibit unique magneto-transport phenomena, including distinct quantum oscillations and zeroth Landau level physics related to the quasi-relativistic dispersion near the nodal lines, conceptually linked to behaviors observed in layered graphene systems~\cite{Rhim2015, Chang2017}. Crucially, the predicted drumhead surface states, a hallmark of NLSMs, have been experimentally resolved using Angle-Resolved Photoemission Spectroscopy (ARPES) in several candidate materials~\cite{Schoop2016, Takane2016, Bannies2021, Regmi2024, Yuan2024, Elius2025}, solidifying their physical realization.
%{\color{blue}*****Ref. [8] is a calculation paper, which cannot support the ARPES experimental evidence.?****}

Over the past decade, research on NLSMs has rapidly evolved from theoretical proposals to material discovery and characterization. This progress has been charted in several valuable review articles that have summarized key developments. These include discussions on the theoretical foundations and topological classification schemes~\cite{Fang_2016}, the prediction and identification of NLSM material candidates using first-principles calculations~\cite{Yu2017}, explorations of unique quantum transport signatures~\cite{Yang31122022}, realizations of nodal-line physics in artificial structures like photonic and phononic crystals~\cite{Park2022}, and comprehensive overviews connecting experimental confirmations with theoretical predictions~\cite{Yang01012018}. While these reviews provide essential perspectives, the field continues to advance rapidly, particularly regarding the diversity of NLSM types and the interplay with interactions.

This review aims to complement the existing literature by providing a focused discussion of various distinct types of NLSMs based on their nodal structures and protecting symmetries. We will delve into their characteristic electromagnetic responses, which serve as powerful probes of their underlying topological nature and band structure. Furthermore, we place particular emphasis on the influence of electron-electron correlation effects in NLSMs, an increasingly critical area exploring how interactions can modify the semimetallic state, potentially driving transitions to novel correlated or symmetry-broken phases. By concentrating on these interconnected aspects, e.g., classification, electromagnetic response, and correlations, we aim to provide an updated perspective on the rich physics harbored within nodal-line semimetals.

\section{Nodal-Line Semimetals (NLSMs)}
\begin{figure*}
\includegraphics[width=0.99\textwidth]{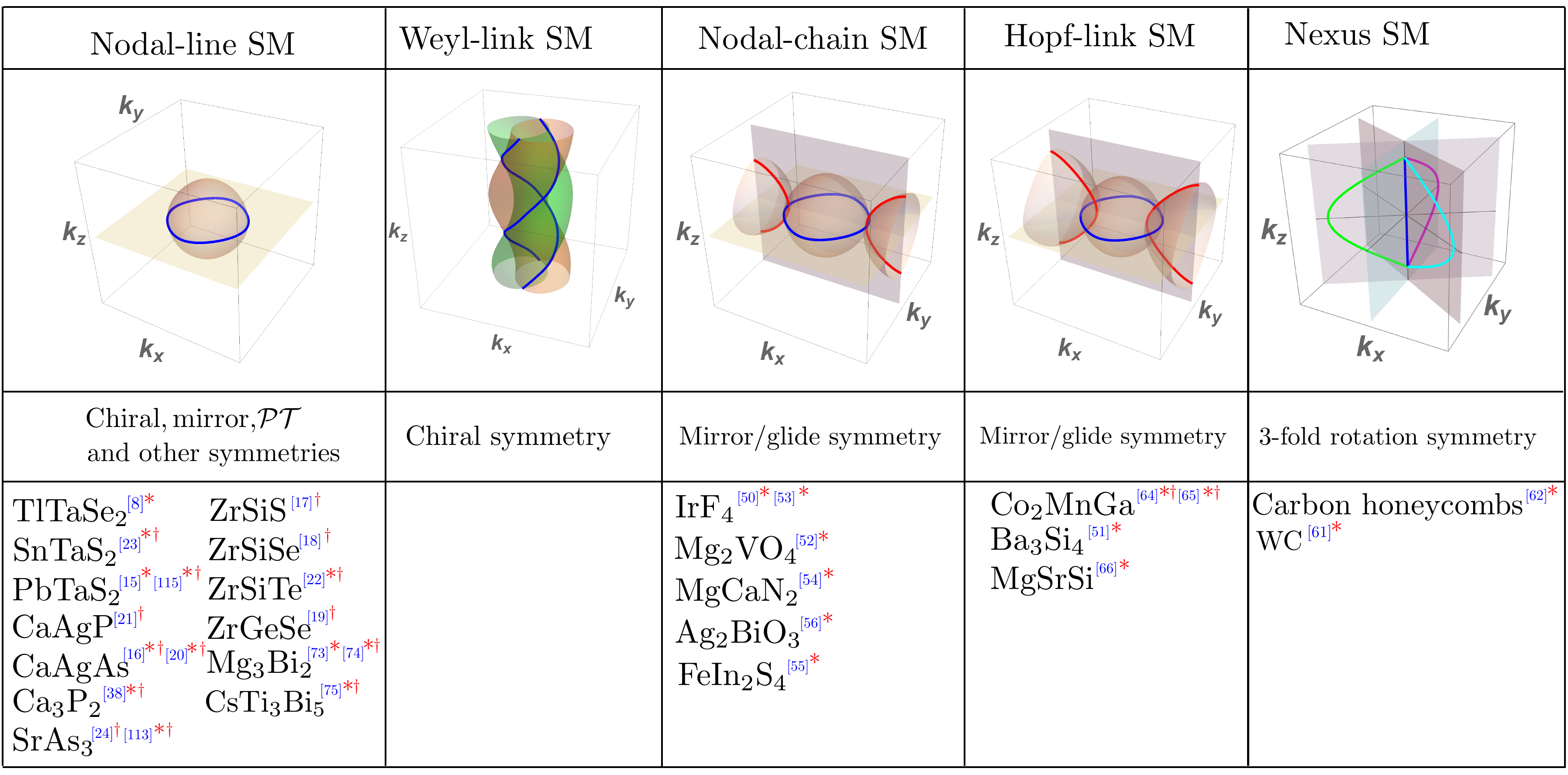}
\caption{We summarize various types of NLSM, highlighting the nodal-line structures, symmetry constraints, and materials candidates.
The theoretical computations are highlighted in $*$ and the experimental observations are marked by $\dagger$, respectively.}
\label{F1}
\end{figure*}

We briefly discuss various types of NLSMs. which are summarized in Fig.~\ref{F1}.

\subsection{Characterizing NLSMs  by symmetries}

A simple NLSM can be described by an effective two-band $k \cdot p$ model
\begin{align}
\mathcal{H}_{\bf k}  = (k_x^2+k^2_y-\mu) \sigma_x +k_z \sigma_y,
\label{Eq:1}
\end{align}
where $\sigma_{\alpha=x,y,z}$ are Pauli matrices and $\mu$ is the chemical potential. The single-particle energy spectrum
is $E^{\pm}_{\bf k} = \pm \sqrt{(k_x^2+k^2_y-\mu) ^2+k_z^2}$, with the nodal line is located on the $k_z=0$ plane, as shown Fig.~\ref{F1}.
The topological invariant associated with this nodal line is given by  $\mathbb{Z}$, which can be computed as the winding number 
for a contour encircling the nodal line
\begin{align}
\nu = \frac{i}{2 \pi} \int_{\mathcal{C}} \langle u^-_{\bf k}| d u^-_{\bf k}\rangle,
\label{Eq:WN}
\end{align}
where $|u^-_{\bf k}\rangle$ is the occupied state with energy $E^-_{\bf k}$.  From Eq.~(\ref{Eq:1}), we can add a mass term $m \sigma_z$ to gapped out the nodal line. 
However, this mass term is forbidden if the system possesses chiral or mirror symmetries. I.e., $\mathcal{S} = \sigma_z$, $\mathcal{S}^\dagger \mathcal{H}_{\bf k} \mathcal{S} = -  \mathcal{H}_{\bf k} $ or   
$\mathcal{M}_z = \sigma_y$, $\mathcal{M}_z^\dagger \mathcal{H}_{\bf k} \mathcal{M}_z =  \mathcal{H}_{\bf \bar {k}} $ with $\bf \bar {k} = (k_x,k_y,-k_z)$. 
We discuss the symmetry-protected nodal lines below.

\subsubsection{Mirror symmetry protected NLSM}
For NLSM protected by the mirror symmetry,
the nodal lines are s on the mirror plane ($k_z=0, \pm \pi$). The nodal lines are formed by the band-degeneracy points. At the mirror plane, the band-degeneracy points possess
 two different mirror eigenvalues, preventing the degeneracy from being lifted. 
The topological invariant associated with a mirror protected nodal line is given by $M\mathbb{Z}$~\cite{chiu2014, Xie2015}. 

\subsubsection{$\mathcal{PT}$ symmetry protected NLSM}
NLSMs can also be protected by the combination of parity $\mathcal{P}$ and time-reversal $\mathcal{T}$ symmetries. 
The $\mathcal{PT}$-protected nodal lines are characterized by a $\mathbb{Z}_2$ topological invariant~\cite{YKim2015, Fang2015, Fang_2016, Zhao2017}.
$\mathcal{PT}$ symmetry enforces the Hamiltonian to be real. One can define the $\mathbb{Z}_2$ topological invariant by finding a one-dimensional line encircling the nodal line or a two-dimensional sphere enclosing it. 
Since these one-dimensional lines or two dimensional spheres  correspond to real and gapped Hamiltonians, the $\mathbb{Z}_2$ topological classification is given by~\cite{hatcher2002algebraic}
\begin{align}
\pi_1 \left (\frac{O(M+N)}{O(M) \oplus O(N)} \right )= \pi_2 \left (\frac{O(M+N)}{O(M) \oplus O(N)} \right )=\mathbb{Z}_2.
\end{align}
The effective Hamiltonian is a four-band model 
\begin{align}
\mathcal{H}_{\bf k} = k_x \sigma_x + k_y \tau_y \sigma_y + k_z \sigma_z + m \tau_x \sigma_x,
\label{HamPT}
\end{align}
where $\sigma_i$ and $\tau_j$ are two sets of Pauli matrices. The nodal line is located on the $k_z=0$ plane.
The energy dispersion from Eq.~(\ref{HamPT}) is $E_{\bf k} = \pm \sqrt{k_z^2 + (\sqrt{k_x^2 +k_y^2 \pm m})^2}$.
By turning the mass term $m$, the nodal line cannot be annihilated by itself. I.e., the radius of the nodal line can only shrink to a point and cannot disappear by varying $m$.
 
\subsubsection{Nonsymmorphic symmetry protected NLSM} 
\label{Sec:nonsymm}
\begin{figure}
\includegraphics[width=0.5\textwidth]{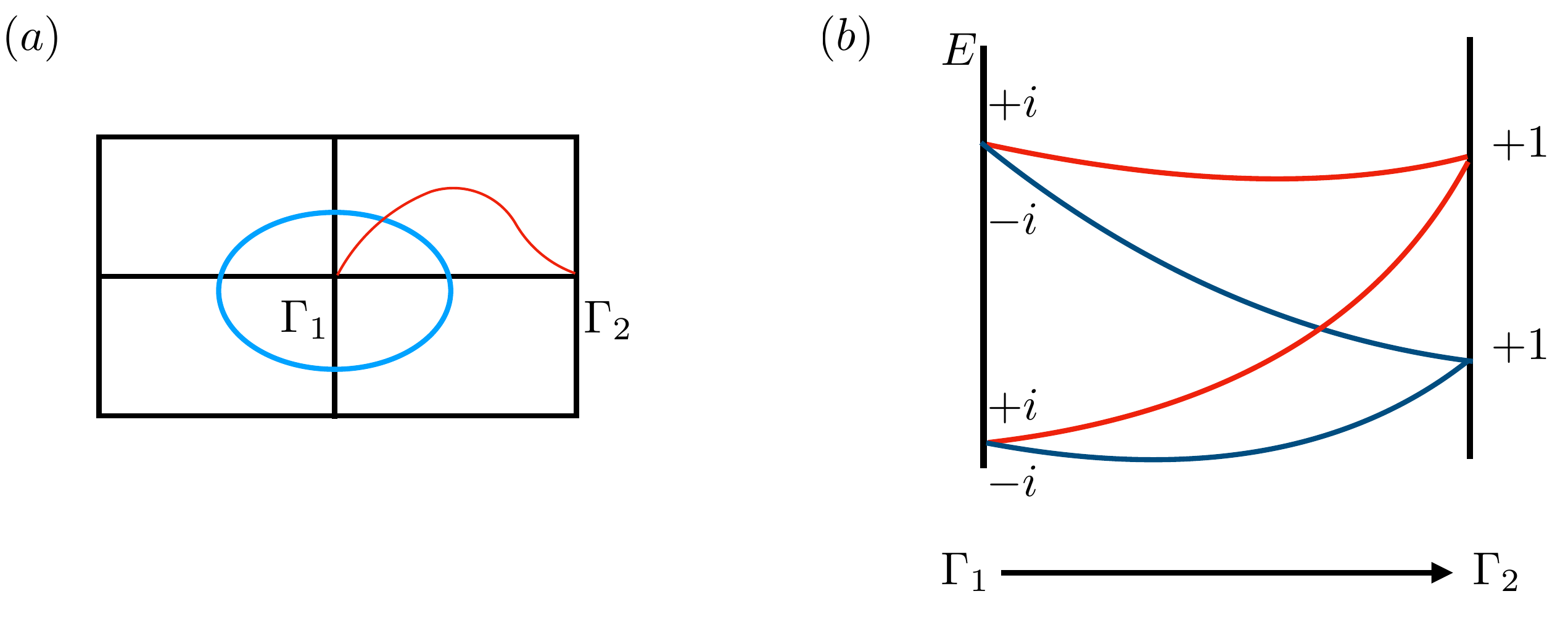}
\caption{NLSMs protected by non-symmorphic symmetries (glide reflection or screw rotation): 
(a) {\bf Glide Mirror Protections:}  Any red path on the glide plane connecting two time-reversal invariant  points $\Gamma_1$ and $\Gamma_2$
will host a degenerate point. Connecting all such degenerate points forms a nodal line (blue circle). (b) The glide eigenvalues $\pm i e^{-i k_x/2}$ at two different time-reversal invariant points $\Gamma_1$ and $\Gamma_2$ are $\pm i$ and $\pm 1$ are switched. I,e,,  $\pm i $ at $\Gamma_1$ becomes $+1$ or $-1$ at $\Gamma_2$. The energy spectrum exhibits an hourglass structure and giving rise to a degenerate point. 
{\bf Screw rotation Protections:} If is an additional inversion symmetry is present, its combination with the screw rotation effectively forms a glide mirror operator. Consequently, a nodal ring can be protected by the combined symmetry, similar to the case of a glide-mirror-protected NLSM.}
\label{F2}
\end{figure}

Nonsymmorphic symmetries such as glide mirror symmetry or screw rotation symmetry can protect the band touching points on the glide mirror plane
or the screw rotation axis~\cite{Fang2015, Young2015, Yang2017, Ahn2018, Takahashi2017, ZWang2019, FURUSAKI2017788}. 
 We provide two examples of the nonsymmorphic symmetry-protected NLSMs below.
 \paragraph{Glide mirror}
 Let us consider the glide reflection $G_z=T_{(1/2,0,1/2)}M_z$. When we square the glide reflection operator, we get
 \begin{align}
 G^2_z =&T_{(1/2,0,1/2)}M_zT_{(1/2,0,1/2)}M_z  \notag\\
 = &T_{(1,0,0)}M_z^2= -T_{(1,0,0)} = - e^{-i k_x}.
 \end{align}
 Here the minus sign is due to the half-integer character of the electron, meaning that the reflection square yuelds $-1$.
 On the glide plane $k_z=0, \pi$, the glide eigenvalues of the bands are $\pm i e^{-i k_x/2}$.
 Suppose the system exhibits time-reversal symmetry (TRS). The time-reversal invariant points (TRIP) on the glide plane must be double degenerate, forming the Kramers' pairs.
 The corresponding  glide eigenvalues at two different TRIPs are $\pm i$ or $\pm 1$ as shown in Fig.~\ref{F2}(a).
 The connectivity of bands along any path on the glide plane connecting two different TRIPs will exhibit hourglass structure, which guarantees a band crossing along the path.
  These  band crossings will form a nodal-ring structure on the glide plane, referred to as a glide mirror protected NLSM.
 
 \paragraph{Screw rotation}
 %{\color{blue} 
 %****
 %The symmetry analysis for the screw rotation is incomplete. The combined symmetries TCy and TCz can guarantee Kramers degeneracy over the whole ky=pi and kz=pi plane, respectively. Whereas, this does not mean the existence of a nodal line along (kx, pi, pi) axis, since the degenerate bands are not gapped on ky=pi and kz=pi off (kx, pi, pi) axis.
 %****
 %}\\
 A similar analysis can be applied to screw rotations combined with  inversion symmetry. Let us consider the two-fold screw rotation $C_x=T_{(1/2,1/2, 0)}S_x$ where $S_x:=(x,y,z)\to (x,-y,-z)$. 
 When Combined with inversion symmetry, the resulting operator is $IC_x=IT_{(1/2,1/2, 0)}S_x = T_{(-1/2,-1/2, 0)}M_x$ where $M_x:=(x,y,z)\to (-x,y,z)$ is the mirror operator.
Notice that this combined operator is similar to a glide mirror operator.
 A similar analysis can now be performed by taking the square of the combined operator.
  \begin{align}
 (IC_x)^2 =&T_{(-1/2,-1/2, 0)}M_x T_{(-1/2,-1/2, 0)}M_x  \notag\\
 = &T_{(0,-1,0)}M_x^2= -T_{(0,-1,0)} = - e^{i k_y}.
 \end{align}
Similar to the glide reflection case with TRS, on the mirror plane at $k_x = 0$ and $\pi$, the eigenvalues of the combined symmetry are $\pm i e^{i k_y/2}$.
At two different TRIPs, the corresponding Kramers' pairs have the eigenvalues $\pi i$ and $\pi 1$ which give rise to the hourglass structure on the mirror plane.
The crossing of the hourglass bands leads to nodal-ring structure as the glide mirror protected NLSM.

 \subsection{Characterizing NLSMs by nodal structures}

\subsubsection{Nodal-link/Weyl-link SMs}
For a two-band model with chiral (sublattice symmetry) $\mathcal{S} \mathcal{H}({\bf k})\mathcal{S}^{-1} = -\mathcal{H}({\bf k})$, where $\mathcal{H}({\bf k}) = f({\bf k})\sigma_x + g ({\bf k})\sigma_y$,
one can find two periodic functions $ f({\bf k})$ and $ g({\bf k})$ such that the nodal lines linked to each other. For simplicity, we consider 
\begin{align}
 f({\bf k}) =& (1 + \cos k_x + \cos k_y), \notag\\
 g({\bf k}) =& [\sin (n k_z) + \cos (n k_z)] \sin k_x\notag\\
 &- [\sin (n k_z) - \cos (n k_z)] \sin k_y,
\end{align}
where $n \in \mathbb{Z}$. This integer $n$ determines the linking number.
For $n=0$, there are two straight nodal lines extending along the $k_z$-direction.
For $n=\pm 1$, these two nodal lines form a helical structure along the $k_z$-direction, which is equivalent to the Hopf link, as shown in Fig.~\ref{F1}.

The linking number associated with the nodal lines can be connected to the Chern-Simons theory.
From Eq. \ref{Eq:WN}, the winding number can be repressed in terms of the Berry connection $\mu = \frac{1}{2 \pi} \oint_{\mathcal{C}} a$,
where the Berry connection is $a = i  \langle u^-_{\bf k}| d u^-_{\bf k}\rangle$. The Chern-Simons $3$-form is defined as
\begin{align}
\theta = \frac{1}{4\pi} \int_{\rm BZ} {\rm Tr} (a \wedge da - \frac{2 i}{ 3} a \wedge a \wedge a).
\end{align}
This Chern-Simons $3$-form describes the electromagnetic response of the three-dimensional topological insulator, where the $\theta$
angle is quantized to be $\pi$ in the topological phase and is zero in the trivial phase due to the time-reversal symmetry.
For nodal-link/Weyl-link SMs, the Chern-Simons $3$-form is related to the linking number
\begin{align}
\frac{1}{4 \pi} \int_{\rm BZ} a \wedge da = \pi \sum_{i,j} \nu_i \nu_j N(\mathcal{L}_i, \mathcal{L}_j),
\end{align}
where $\nu_i$ is the vorticity associated with the nodal lines, and $ N(\mathcal{L}_i, \mathcal{L}_j)$ is the Gauss linking integral,
\begin{align}
 N(\mathcal{L}_i, \mathcal{L}_j) = \frac{1}{4 \pi} \oint_{\mathcal{L}_i} dx^i \oint_{\mathcal{L}_j} dy^j  \epsilon_{ijk} \frac{(x-y)^k}{|x-y|^3},
\end{align}
which determines the number of times two loops $\mathcal{L}_i$ and $\mathcal{L}_j$ linked with each other. 
Similar constructions of Nodal-link/Weyl-link SMs are discussed in Refs.~\cite{Chang2017, Chen2017, Yan2017, Ezawa2017}.

\subsubsection{Nodal-chain SMs}
%{\color{blue}
%****
%The nodal chain not only can be protected by nonsymmorphic symmetries, but also can be guaranteed by symmorphic symmetries, i.e. two mirrors perpendicular to each other. Such a nodal chain has been discussed in several works, such as Phys. Rev. Lett. 119, 036401 (2017), Phys. Rev. Lett. 119, 156401 (2017), Nat. Phys. 14, 461 (2018), Phys. Rev. Lett. 120, 106403 (2018), npj Quantum Mater. 7, 52 (2022), etc.
%****}\\
The nodal-chain SM was first introduced in Ref.~\cite{Bzdu2016}. In this phase,
the nodal lines form a chain-like structure where
all the nodal lines are connected to each others. The first criterion for hosting a nodal-chain SM phase is the existence of the nonsymmorphic symmetry-protected NLSMs (specifically, glide mirror), as discussed in Sec.~\ref{Sec:nonsymm}.
The system also needs to possess two inequivalent glide planes, with time-reversal invariant points (TRIPs) located at the intersections of these planes.
The final criterion is that the intersection of the nodal loops forming the chain much belong to the two-dimensional representations at the TRIPs.
The chain configuration is illustrated in Fig.~\ref{F1}. Similar nodal-chain SMs are also discussed in Refs.~\cite{Cai2018, Zhang2020, Shang2023, Wu_2023, Liu2024, Fu2018}.
In addition to be protected by nonsymmorphic symmetries, nodal-chain SMs can also be guaranteed by certain symmorphic symmetries, specifically those composed by two perpendicular  mirrors.
These types of nodal-chain SMs are discussed in Refs.~\cite{Yu2017, GChang2017, Ahn2018, Yan2018}.

\subsubsection{Nexus fermion, triplet points, and their nodal networks}
Triplet degenerate points that can be visualized as the terminal points of several nodal lines are referred to as the Nexus points~\cite{Heikkila_2015, Hyart2016, Chang_2017}.
These Nexus points can be protected by three-fold rotation symmetry, and the effective $k\cdot p$ model can be written as
\begin{widetext}
\begin{align}
\mathcal{H}(\bf k) = \left(\begin{array}{ccc}A_1 (k_x^2 +k_y^2) + B_1 \cos k_z + C_1 & \alpha (k_x+i k_y) \sin k_z + \beta (k_x-ik_y)^2 & D(k_x-i k_y) \\\alpha (k_x-i k_y) \sin k_z + \beta (k_x+ik_y)^2 & A_1 (k_x^2 +k_y^2) + B_1 \cos k_z + C_1 & -D(k_x+i k_y) \\D(k_x+i k_y) & D(k_x-i k_y) & A_2 (k_x^2 +k_y^2) + B_2 \cos k_z + C_2\end{array}\right).
\end{align}
\end{widetext}
For $\alpha=0$, the Hamiltonian exhibits a six-fold rotation symmetry. 
For  $\alpha \neq 0$, the six-fold rotation symmetry is broken down to a three-fold rotation symmetry. 
The triplet point (three-fold degeneracy) is located on the rotation axis.
Besides the  three-fold rotation symmetry, there are three additional mirror planes, each hosting nodal lines. 
These nodal lines merge at the intersection of these mirror planes, i.e., at the rotation axis.
The merging point is the Nexus point, where the first two-fold degenerate nodal-line (formed by the first and the second bands) meets
the second two-fold degenerate nodal-line (formed by the second and the third bands). Therefore, the merging point is a three-fold degeneracy point.

The first two-fold degenerate nodal-line and the second two-fold degenerate nodal-line, together with their merging points (Nexus points) arrange themselves as a chain-like structure,
referred to as the Nexus network~\cite{Chen2018}.  It should be noted that this network structure differs from nodal-chain SMs. The merging points of the  nodal-chain SMs are still two-fold degenerate, unlike the Nexus network
where the merging points are three-fold degenerate. 

We can also consider another variance of the nodal network by further breaking the rotation symmetry. 
For the triplet point that is protect by $C_4$ screw rotations~\cite{Xie2019}, breaking the
$C_4$ screw rotations can lead to the splitting of the triple point into two-fold degenerate points.
The first two-fold degenerate nodal-line and the second two-fold degenerate nodal-line do not merge and become two individual nodal loops on their corresponding mirror planes.
These nodal loops can form non-trivial link (Hopf link) and thus form Hopf-link networks~\cite{Chang2017, Belopolski2019, Belopolski2022, Cai2018, Xie2019, Lian2019}.

\subsubsection{Other structures}
In addition to the Nodal-link/Weyl-link SMs, nodal lines can also self-linked, giving rise to nodal-knot SMs~\cite{Bi2017}.
Furthermore, nodal lines can connect to form a box-like nodal structures, as discussed in Ref.~\cite{Sheng2017}.
Dirac nodal-chain SMs, in which the nodal lines possess a fourfold degeneracy, were proposed in Ref.~\cite{Wang2017_2}.
In the presence of multiple mirror symmetries,  nodal lines can cross at the intersection of  the mirror planes,  leading to a NLSM referred to as crossing-line-node SMs~\cite{Kobayashi2017}.

\subsection{Characterizing NLSMs by dispersions}
\subsubsection{Type-I/II/III NLSMs}

\begin{figure}
\includegraphics[width=0.5\textwidth]{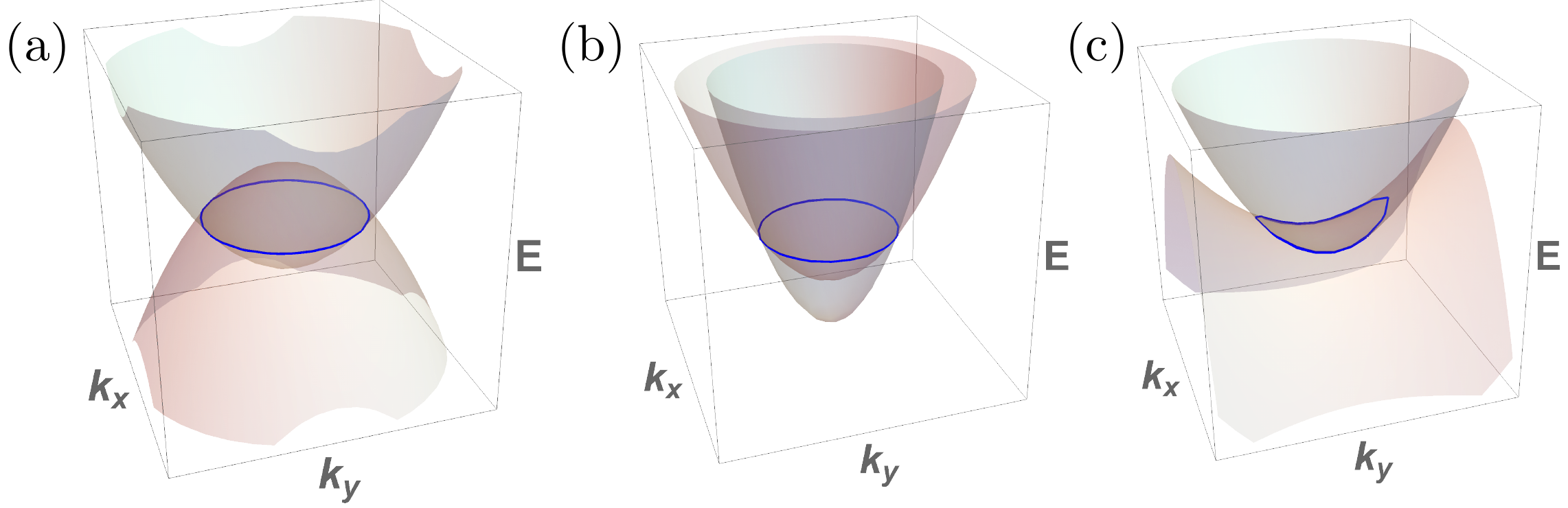}
\caption{Classification of energy dispersions near the nodal line: (a) Type I, (b) Type II, and (c) Type III. }
\label{F3}
\end{figure}

Nodal lines can be classified into three types based on their differing dispersions~\cite{Gao2018}.
This classification can be obtained using a simple two-band model at the $\Gamma$ point 
\begin{align}
\mathcal{H}_{\bf k} = &\frac{1}{2}\left( (A_1+A_2) k_x^2 +(B_1+B_2)k_y^2 +\Delta \right)\mathbb{I}  \notag \\
&+ \frac{1}{2}\left( (A_1-A_2) k_x^2 +(B_1-B_2)k_y^2 -\Delta \right)\sigma_z+ C k_z \sigma_y,
\end{align}
where $A_{1}, A_{2}, B_{1},B_{2},C$ and $\Delta$ are tuning parameters that can be determined from the {\it ab initio} calculation.
For simplicity, we set $C=1$. 
\begin{itemize}
\item Type-I: When $A_1,B_1> 0$, $A_2, B_2 <0$, the NLSM is classified as type-I, as shown in Fig.~\ref{F3}(a).
\item Type-II: When $A_1,A_2, B_1, B_2> 0$ or $A_1,A_2, B_1, B_2< 0$ , the NLSM is classified as type-II, as shown in Fig.~\ref{F3}(b).
\item Type-III: When $A_1,A_2, B_1>0$, $B_2 <0$, or $B_1,B_2, A_2<0$, $A_1 >0$, the NLSM is classified as type-III, as shown in Fig.~\ref{F3}(c).
\end{itemize}
These different dispersions can strongly influence the transport properties, as we will discuss the Landau level spectrum in Sec.~\ref{LL}.
Most of the NLSMs are belong to type-I. 
The material K$_4$P$_3$ was firstly proposed as a candidate for a type-II NLSM in Ref.~\cite{Li2017}, and 
Mg$_3$Bi$_2$ was later also shown to host this state~\cite{Zhang2017, TRChang2019}.
A candidate material for a type-III NLSM is CsTi$_3$Bi$_5$~\cite{JYang2023}.
For type-III nodal line, the nontrivial Fermi surface can lead to unconventional magnetic responses, such as the zero-field magnetic breakdown and the momentum-space Klein tunneling~\cite{Zhang2018}. These features are distinct from those of type-I nodal lines.

\subsubsection{Higher-order dispersions}
Quadratic and cubic nodal lines can also be stabilized by multiple crystalline symmetries~\cite{Yu2019}.
Let us consider a Hamiltonian with a threefold rotation $C_{3Z}$ and a combined symmetry $\mathcal{T}M_z$,
where $\mathcal{T}$ is the time-reversal symmetry and $M_z$ is the mirror symmetry on the $x-y$ plane. Here, the time-reversal symmetry satisfies $\mathcal{T}^2=-1$.
The effective $\bf{k} \cdot \bf{p}$ model around the nodal line can be expressed as 
\begin{align}
H_{\rm{eff}} (\bf{k}+\bf{q}) = f(\bf{q}) + g(\bf{q}) \sigma_+ +g^*(\bf{q}) \sigma_- +h(\bf{q}) \sigma_z,
\end{align}
where $\bf{q}$ is a small wavevector around a point $\bf{k}$ on the nodal line, and $g(0) = h(0) = 0$.
The symmetries constraints on the effective Hamiltonian around the nodal line are
\begin{align}
&C_{3z}H_{\rm{eff}} (\bf{q})C_{3z}^{-1} = H_{\rm{eff}} (R_{3z}\bf{q}), \notag\\
&(\mathcal{T}M_z)H_{\rm{eff}} (\bf{q})(\mathcal{T}M_z)^{-1}=H_{\rm{eff}} (-\bf{q}).
\end{align}
Now, let us suppose the nodal line is along $k_z$ direction. The wavevector $\bf{q}$ is in the transverse plane $(q_x,q_y)$ and can be expressed as $q_{\pm} = q_x \pm q_y$.
The symmetry operators can be represented as $C_{3z} = e^{i \frac{2 \pi }{3} \frac{\sigma_z}{2}}$ and $\mathcal{T}M_z = - i \sigma_x \mathcal{K}$ with $ \mathcal{K}$ being the complex conjugation operator.
The combined operator $\mathcal{T}M_z $ restricts $g(\bf{q})$ and $h(\bf{q})$ to be even functions, and hence the dispersion cannot be linear around the nodal line.
The threefold rotation symmetry $C_{3z}$ has the property $C_{3z} \sigma_{pm}C_{3z}^{-1} = e^{\pm i 2\pi/3} \sigma_{\pm}$. The corresponding leading order term is quadratic, and the effective Hamiltonian is
\begin{align}
H_{\rm{eff}} (\bf{q}) = \alpha q_-^2 \sigma_+ +\alpha^* q_+^2 \sigma_-.
\end{align}

For a cubic dispersion, a similar analysis can be applied with a sixfold rotation $C_{6z}$ and a mirror symmetry $M_x$.
The standard basis function~\cite{bradley2010mathematical} considered with respect to these symmetries is $\{ |\frac{3}{2}, \frac{3}{2} \rangle, | \frac{3}{2}, -\frac{3}{2}\rangle \}$.
The corresponding basis representations of the symmetry operators are $C_{6z} = i \sigma_z$ and $M_x = \sigma_x$.
The symmetries constrains on the effective Hamiltonian around the nodal line are
\begin{align}
&C_{6z}H_{\rm{eff}} (\bf{q})C_{6z}^{-1} = H_{\rm{eff}} (R_{6z}\bf{q}), \notag\\
&M_xH_{\rm{eff}} (\bf{q})M_x^{-1}=H_{\rm{eff}} (\bar{\bf{q}}).
\end{align}
Here $\bar{\bf{q}}=(-q_x, q_y)$. The sixfold rotation symmetry forbids the linear terms in $g(\bf{q})$ and $h(\bf{q})$. Using the fact that $(R_{6z}q_{\pm})^3 = - q_{\pm}^3$,
the corresponding leading order is cubic, and the effective Hamiltonian is
\begin{align}
H_{\rm{eff}} (\bf{q}) = i (a q_-^3 +b q_+^3) \sigma_+- i (a q_+^3 +b q_-^3) \sigma_-,
\end{align}
where $(a,b)$ are real functions depend on $k_z$.

\subsection{Characterizing NLSMs by textures}

\subsubsection{Pseudospin Vortex Ring SMs}
Beside symmetry-protected nodal lines, 
one can also consider the nodal line as vortex line, where the pseudospin texture around the nodal line forms a toroidal vector field around its axis~\cite{Lim2017}.
The effective two-band Hamiltonian is given by
\begin{align}
\mathcal{H}_{\bf k} =& -\frac{1}{m_z} k_x k_z \sigma_x -\frac{1}{m_z} k_y k_z \sigma_y \notag\\
&+ \left( \frac{1}{m_r} (k_x^2+k_y^2 -k_z^2-k_0^2)\right),
\end{align}
 where the nodal line forms a ring on the $k_z=0$ plane, $k_0$ sets the radius of the ring, and $m_{r/z}$ sets the Fermi velocities.
 The pseudospin texture can be understood from the following procedure. First, for a fixed $k_y$ plane that intersects with the nodal ring, there are two Dirac points.
 Each of the Dirac point can be expressed as $\mathcal{H}_{\bf k} = -k_z \sigma_x + k_x \sigma_z$. The pseudospin winds a $2\pi$ angle along a trajectory that encircles the Dirac point.
 Hence this nodal ring can be viewed as a vortex ring.

%\subsubsection{Bulk-boundary correspondence---Drumhead surface states}
%\subsubsection{Material realizations}
 
%Refs.~\cite{Burkov2011, Zhao2013, Matsuura_2013, Fang2015, Fang_2016, Zhao2017, Lim2017, Sun_2017, Wang2017, Ratnadwip2017, Pan2018,Guo2019, Hirose2020,Okamoto2020}

%materials~\cite{Bian2016,Stuart2022,Gao2023,Jeon2023}

\subsubsection{Nodal lines with non-Abelian charges}
For multi-band systems, there can be multiple nodal lines formed from the crossings of difference bands.
These nodal lines can carry non-Abelian charges, as discussed in Ref.~\cite{QuanShengWu2019}.
Let us take a three-band model  as an example. The flattened Hamiltonian in the eigen-basis can be expressed as $H_3(\bf{k}) = \sum_{j=1}^3 \epsilon_j |u_{\bf{k}}^j \rangle \langle  u_{\bf{k}}^j  | $ with $ \epsilon_j=j$.
This flattened Hamiltonian can be thought of as a manifold encoded by a frame $\{  |u_{\bf{k}}^j \rangle \}$ of orthonormal $3$-component vectors, modulo $|u_{\bf{k}}^j \rangle \to - |u_{\bf{k}}^j \rangle$.
This manifold is $M_3 = SO(3)/D(2)$, where $D_2$ is the three-dimensional dihedral point group that contains the $\pi$ rotations about three axes. I.e., the $\pi$ rotation corresponds to the transformation $|u_{\bf{k}}^j \rangle \to - |u_{\bf{k}}^j \rangle$.
Similar to how we classify the topological charge of conventional nodal lines, we need to find the close path that encircles the nodal lines. The associated topological charges is characterized by the quaternion group
\begin{align}
\pi_1 (M_3) = \mathbb{Q} = \{ \pm1, \pm i, \pm j , \pm k\},
\end{align}
where $i^2=j^2 = k^2=-1$ and $i$, $j$, and $k$ are anticommuting. I.e., the topological charges are non-Abelian. 
These non-Abelian NLSMs have the similar structure as the biaxial nematic liquid
crystals, which can host non-Abelian disclination lines~\cite{Madsen2004, Kleman_1977, Mermin1979}.
The texture near the  disclination lines in  biaxial nematic liquid crystals is analogous to  the  linear-polarized electric field near the non-Abelian nodal lines  in the photonic metamaterials~\cite{Yang2020}.

%{\color{blue}
%****
%The type-II nodal line was firstly proposed in Phys. Rev. B 96, 081106(R) (2017), which should be addressed. For the type-III nodal line, the nontrivial Fermi surface can lead to unconventional magnetic responses, such as the zero-field magnetic breakdown, the momentum-space Klein tunneling, etc, as shown in Phys. Rev. B 97, 125143 (2018). This feature is distinct from that of type-I nodal lines. It is recommended to discuss it in the review.
%****
%}

\section{Electromagnetic response}
The electromagnetic response in these NLSMs is very rich due to their anisotropy and tunability. 
Similar to other semimetals, such as graphene and Weyl semimetals, the transport properties under an external magnetic field are strongly related to the topological nature of their nodal structures. 
Furthermore, the optical conductivity can be sensitive to both their band dispersions and nodal structures. 
As discussed in the previous section, nodal structures are related to crystal symmetry and can be broken by applying external strain or fields. 
Additionally, the dispersion can be highly anisotropic. All these properties suggest a huge potential for applications stemming from their electromagnetic response.
\subsection{Landau level}
\label{LL}
For NLSMs in three dimensions, consider a given two-dimensional slice embedded in the 3D Brillouin zone. There can be several gapless points where this two-dimensional slice intersects with the nodal lines. 
These gapless points within the two-dimensional slice are analogous to the Dirac cones found in a graphene sheet. When an external magnetic field is applied perpendicular to graphene, a zeroth Landau level appears at zero energy. For NLSMs subjected to an external magnetic field, one can visualize the perpendicular 2D slices as graphene-like systems with Dirac cones, where zeroth Landau levels are associated with each of these graphene-like slices. These zeroth Landau levels are linked to forming a dispersionless band~\cite{Rhim2015, Chang2017}. These flat bands strongly depend on how the two-dimensional slices (perpendicular to the field) intersect with the nodal lines. In principle, by applying Landau level spectroscopy with varying field directions, one can map out the nodal-line structure of an NLSM.
Besides the Landau level spectrum, if the NLSM hosts toroidal Fermi surfaces around the nodal rings, the system can exhibit three-dimensional quantized Hall conductivities under a magnetic field~\cite{Halperin_1987, Koshino2002, Bernevig2007, Mullen2015}.

%\subsection{transport}
%Packet-hole

\subsection{Optical conductivity/permittivity, and hyperbolic polariton}
In Weyl and Dirac SMs, the low-energy optical spectroscopy strongly depends on their anisotropic dispersion~\cite{Carbotte2016},
and the temperature dependence is strong due to the quadratic nature of the density of states near the nodal points~\cite{Singh2019, Singh2021}.
The topological features of these Weyl/Dirac SMs are also reflected in their optical properties~\cite{Morimoto2016, Morimoto2018, Juan2017}. 
It is therefore important to Investigate the optical properties in NLSMs, especially the relationship between  the polariton and the electronic structure of the nodal lines.
Due to the highly anisotropic electronic structure in NLSMs, i.e., the quasipartical dispersion is parabolic in the nodal-line plane and is linear in the transverse direction, the dielectric permittivity tensor is highly
anisotropic ~\cite{Singh2023}.
This property can be analyzed by directly computing the optical conductivity from the Kubo formula

\begin{align}
\sigma_{\alpha \beta} (\omega) = \frac{i \hbar}{V} \sum_{m,n,\sigma} \frac{f_{n\sigma}-f_{m \sigma}}{E_{m \sigma}- E_{n \sigma}} \frac{ \langle \psi_{n\sigma} |J_\alpha| \psi_{m \sigma} \rangle \langle \psi_{m\sigma} |J_{\beta}| \psi_{n\sigma} \rangle}{\hbar (\omega + i \Gamma) +E_{n \sigma} - E_{m \sigma}},
\end{align}
where $\alpha, \beta = x,y,z$, $\sigma = \uparrow, \downarrow$, $m, n$ are the band labels, $f_{m \sigma} = (1+ e^{(E_{n \sigma} - \mu)/k_BT})^{-1}$, $T$ is the temperature,
$\mu$ is the chemical potential, $J_{\alpha} = \hbar^{-1} \partial_{k_\alpha} \mathcal{H}_{\bf k}$ is the current operator,
and $\Gamma$ is the phenomenological decay term.

We can convert the optical conductivity to the dielectric permittivity tensor using  $\epsilon(\omega)_{\alpha\beta} = \epsilon_b \delta_{\alpha\beta} + i \sigma_{\alpha\beta} / (\omega \epsilon_0)$,
where $\epsilon_b$ is the background permittivity. The dielectric permittivity tensor can be highly anisotropic, meaning $\epsilon_{xx} =\epsilon_{yy} \neq \epsilon_{zz} $.
In certain parameter regions, the elements of the dielectric permittivity tensor can have opposite signs $\epsilon_{xx}>0$ and $\epsilon_{zz}<0$.
The propagation of  EM waves in the media is governed by the Maxwell's equations ${\bf n} ({\bf n}  \cdot {\bf E} ) -n^2 {\bf E} + {\bf \epsilon } {\bf E} =0 $,
with ${\bf n} = {\bf q} c/\omega $. The solution for the EM wave has two linearly polarized modes, which are referred to as ordinary and extraordinary waves.
The refractive indices associated with these modes are
\begin{align}
n_o^2 = \epsilon_{xx}, \quad n_e^2 = \frac{\epsilon_{xx} \epsilon_zz}{\epsilon_{xx} \sin^2 \theta + \epsilon_{zz} \cos^2 \theta},
\end{align}
where $\theta = \cos^{-1} (n_z/|{\bf  n}|)$. For the extraordinary mode, the opposite signs between $\epsilon_xx$ and $\epsilon_xx$ lead to the hyperbolic surface
$n_x^2/ \epsilon_{zz}+ n_z^2/\epsilon_{xx}= 1$. Here, we set $n_y=0$ for simplicity. This mode is referred to as the hyperbolic polariton in NLSMs~\cite{Singh2023}.

\section{Correlations}

\begin{figure}
\includegraphics[width=0.5\textwidth]{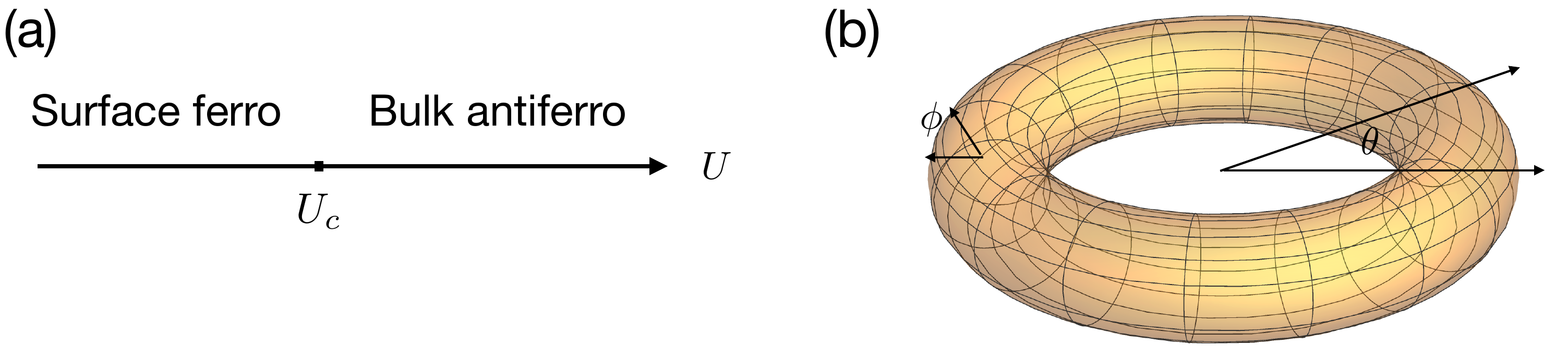}
\caption{(a) The phase diagram of magnetic orders as a function of the Hubbard interaction $U$. Below a threshold $U_c$, the surface develops ferromagnetic order. Above the threshold $U_c$, the bulk develops antiferromagnetic order.
The torus-like Fermi surface of the NLSMs, which can be parameterized by two angles $(\theta,\phi)$.}
\label{F4}
\end{figure}

One of the topological properties of  NLSMs is the existence of drumhead  surface states. These surface states are typically nearly flat. In the presence of chiral symmetry, 
these surface states will be exactly pinned at zero energy. These (nearly) flat bands on the surface can lead to a divergence of density of states (DOS). As a results, the correlation effects are strong.
Due to the strong correlations on the surfaces, the symmetry breaking orders. such as magnetism and superconductivity, are expected. On the other hand, the nodal-line structures in the bulk
result in multiple Fermi surfaces or a small DOS, depending on the dispersions and  chemical potential near the nodal lines. Theoretically investigations of the correlation effects in NLSMs are usually based on simple models.
Several theoretical proposals suggest surface topological orders and unconventional superconductivities can exist in these NLSMs. In the following, we review both the theoretical investigations and the experimental observations of  correlation effects in  NLSMs.

\subsection{magnetization}
The drumhead surface states can produce a large DOS on the surface and lead to several instabilities, In terms of  magnetism, the Stoner ferromagnetic order is preferred on the surface~\cite{Lothman2017, Liu2017, Chen2019}.
This can be understood from the mean-field approximation of the Hubbard interaction $H_{\rm int.}= U \sum_{i} n_{i \uparrow} \langle n_{i\downarrow}\rangle+n_{i \downarrow} \langle n_{i\uparrow}\rangle-\langle n_{i \uparrow}\rangle \langle n_{i\downarrow}\rangle$. The ground state energy is minimized when $ \langle n_{i\uparrow}\rangle \neq  \langle n_{i\downarrow}\rangle$, which leads to ferromagnetism.
Theoretically, the surface ferromagnetic order can further gap out the surface states and give rise to a surface Chern insulator~\cite{Chen2019}. This surface Chern insulating state hosts chiral edge modes that act as hinge states in these NLSMs.

From the bulk perspective, if the nodal lines are pinned at the Fermi level, the DOS vanishes and the electron-electron interactions are irrelevant. However, strong interactions can still drive NLSMs into symmetry-breaking phases.
In Ref.~\cite{Roy2017}, the author considered the onsite ($U$) and the nearest-neighbor ($V$) repulsive interactions, and found that the bulk magnetic orders could be either antiferromagnetic ($U$ dominated) or ferromagnetic ($V$ dominated). As shown in Fig.~\ref{F4}(a), for a small onsite interaction $U$, a surface ferromagnetic order first develops. The bulk antiferromagnetic order starts to form for a larger onsite $U$. The surface magnetic order can introduce a proximity effect in the bulk~\cite{Roy2017}. 
Although a given surface is ferromagnetically ordered, the opposite surfaces have ferromagnetic orders pointing in opposite directions, and the entire system in the slab geometry is still antiferromagnetic. 
Another important feature of the surface magnetic orders is the existence of surface magnons. In Ref.~\cite{Alassaf2024}, it is shown that the magnon spectra can be tuned by the thickness of the slab geometry. Besides creating surface magnetic orders,  bulk interactions can also drive systems into a nodal-line semimetallic phase. 
In particular, the Hund coupling between the local moments and itinerant electron spins can lead to nodal-line semimetallic phases~\cite{Geilhufe2019, Yang_2024}.

\subsection{superconductivity}
Similar to surface magnetic orders, the large DOS on the surface can also lead to superconductivity~ \cite{Lothman2017, Kopnin2011}. 
Surface superconductivity has been experimentally reported for  CaAg$_{1-x}$Pd$_x$P~\cite{Yano2023}.
For bulk superconductivity, theoretical models often begin with the torus-like Fermi surface~\cite{Sur_2016, Nandkishore2016, Wang_2017, Shapourian2018, Ahn2020, Lau2021, Wu2023, Wang_2022} [see Fig.~\ref{F4}(b)].
For this torus-like Fermi surface, the superconducting order parameter $\Delta(\theta, \phi)=\sum_{l_1}e^{i \theta l_1}( \Delta_{l_1,l_2} \cos(\phi l_2) + \bar{\Delta}_{l_1,l_2} \sin (\phi l_2))$ can be parametrized by two angles $(\theta, \phi)$ and expressed in the angular momentum basis $(l_1,l_2)$. The Fermionic statistics restrict the pairing function $\Delta(\bm{p}) = - \Delta(- \bm{p})$. The energetically favored pairing function is the fully gapped one, with $l_1$ being odd and $l_2=0$.
This paring function corresponds to chiral superconductivity, which breaks the time-reversal symmetry~\cite{Nandkishore2016}.
Fragile and  higher-order topological superconductors, as well as nodal superconductivity, have also been 
proposed by including the crystalline symmetry together with the torus-like Fermi surface~\cite{Shapourian2018, Ahn2020, Wang_2022, Wu2023}.

Theoretically, the topological properties of  NLSMs in their normal states can carry over to their superconducting states. However, most experimental observations have not yet confirmed the existence of topological superconductivity~\cite{Muechler2019, Cheng2020,Li2022, Gao2020, Ikeda2022, Ikenobe2023}.
Nevertheless, unconventional superconductivity has been reported, including:
\begin{itemize}
\item Multi-gap structures in CaSb$_2$~\cite{Duan2022, Takahashi2021}.
\item Time-reversal-symmetry breaking of the superconducting states in LaNiGa$_2$~ \cite{Badger2022}, LaNiSi, LaPtSi, and LaPtGe~\cite{Shang2022}.
\item Anisotropic gap structure in SnTaS$_2$~\cite{Singh_2022} and In$_x$TaS$_2$~\cite{Li2020}.
\item $s+ip$ superconductivity in $T$RuSi ($T$ being transition metals)~\cite{Tian2022}.
\item Coexistence of topological surface states and bulk superconductivity in PbTaSe$_2$~\cite{Guan2016}.
 \end{itemize}

\subsection{Kondo effects}
The Kondo effect describes the interaction of free conduction electrons with magnetic moments in materials, which can lead to a strong renormalization of effective mass and band dispersion. 
These magnetic moments are usually formed by localized $f$-electrons. If these $f$-electrons are arranged into a lattice structure (i.e., a Kondo lattice), the corresponding dispersion becomes flat. The effective Hamiltonian is the Anderson lattice model
\begin{equation}\label{}
H=\sum_{i,j,\sigma,\sigma'} \Psi_{i\sigma}^\dagger \mathcal{H}_{ij,\sigma\sigma'}\Psi_{j\sigma'}+U\sum_i n_{if \uparrow} n_{if \downarrow},
\end{equation}
where
%\begin{widetext}
\begin{align}
%H=\sum_{i,j,\sigma, \sigma'} \left(\begin{array}{cc}c^\dagger_{i,\sigma}, & f^\dagger_{i,\sigma}\end{array}\right)
\mathcal{H}_{ij,\sigma\sigma'}=\left(\begin{array}{cc} (
-t^c_{i,j}-\mu^c\delta_{ij})\delta_{\sigma \sigma '} &
{V}_{\sigma \sigma'}({\bf R}_i-{\bf R}_j) \\ {V}_{\sigma
\sigma'}({\bf R}_i-{\bf R}_j) & (
-t^f_{i,j}-\mu^f\delta_{ij})\delta_{\sigma \sigma
'}\end{array}\right).
%\left(\begin{array}{c}c_{j\sigma'} \\f_{j\sigma'}\end{array}\right)
%+U\sum_i n_{if \uparrow} n_{if \downarrow}.
\end{align}
Here $t^{c/f}_{i,j}$ is the hopping terms for conduction/$f$-electrons, $\mu^{c/u}$ is the chemical potential of conduction/$f$-electrons,
$ {V}_{\sigma \sigma'}({\bf R}_i-{\bf R}_j)$ represents the hybridization between conduction and $f$-electrons,
 $\Psi_{i\sigma}^\dagger=(c^\dagger_{i\sigma}$, $f^\dagger_{i\sigma})$ with $c^\dagger_{i\sigma}$ and $f^\dagger_{i\sigma}$ are the creation operators for conduction and f-electrons.

One can derive a simple model of the above Hamiltonian under the self-consistant mean-field approach in the large $U$ limit~\cite{Chang2018}.
The mean-field Hamiltonian takes the form
$H=\sum_{\bf k}\Psi^\dagger({\bf k})\mathcal{H}({\bf k})\Psi({\bf k})+\mathcal{N}_s\lambda(|b|^2-Q)$ with
\begin{align}
\mathcal{H}({\bf k})=&\left(\begin{array}{cc}\epsilon_c({\bf k})-\mu & \sum_j V_j \sigma_j \sin k_j \\
\sum_j V_j \sigma_j \sin k_j  & \epsilon_f({\bf k})+\lambda\end{array}\right)  \notag \\
&+ \left(\begin{array}{cc} 0 & W_0 +i \vec{W}\cdot \vec{\sigma }
 \\
W_0 -i \vec{W}\cdot \vec{\sigma } & 0 \end{array}\right).
\label{Ham}
\end{align}
Here $V_i=v_ib$ and $W_{i}={w}_ib$ are the renormalized hybridization terms with $b$ being the slave boson projection amplitude.
The $f$-electron hopping amplitude is renormalized as $\tilde{t}^f=b^2 t^f$. For simplicity, we take the dispersion of the conduction electrons as $\epsilon_c({\bf k})=-2\sum_i t_i \cos k_i $
and $\epsilon_f({\bf k})=\alpha \epsilon_c({\bf k})$.
Under the mean-field constraint  $Q=n_f+b^2$, we introduce the constraint field $\lambda$.
Here $Q$ is the local conserved charge associated with the slave boson approach in the infinite $U$ limit,
and it is taken to be $Q=1$. $\mathcal{N}_s$ is the total number of sites.
When $\vec{W} = 0$ with non-vanishing $W_0$, the system exhibit two nodal rings protected by the chiral symmetry. The dispersion near the nodal rings are strongly renormalized by the f-electrons.
 
Beyond this simple model, other topological nodal-line "Kondo" semimetals have been reported including
Ce$_3$Pd$_3$Bi$_4$~\cite{Cao2020},  
 CePt$_2$Si$_2$~\cite{Ma2023}, 
 Ce$_2$Au$_3$In$_5$~\cite{Chen_Si_2022},
 and antiferromagnetic CeCo$_2$P$_2$~\cite{liu_2024}.

%\section{Experiments}
%\subsection{ARPES}
%\subsection{Transport}
%Refs.~\cite{Yu2015, Bian2016,Chan2016}

\section{conclusion and outlook}
We review different types of NLSMs whose nodal structures depend on their underlying crystalline symmetries. 
The electromagnetic responses of these NLSMs are highly sensitive to their nodal structures, providing platforms for transport and optical applications. 
The correlation effects in these materials are rich, giving rise to magnetic orders, unconventional superconductivity, and heavy fermion physics. 
Furthermore, several promising future directions for NLSMs could lead to interesting material properties and potential applications:

\begin{itemize}
\item Heterostructures and Interfacial Physics:
A heterostructure is an engineered material created by layering different materials together. The interface between an NLSM and another material (such as a superconductor, a magnet, or a conventional insulator) is a particularly rich platform for discovering new physics. Manipulation of Drumhead States: NLSMs host unique surface states known as "drumhead states." The energy dispersion of these states is highly sensitive to the boundary conditions at the material's surface. By creating a heterostructure, the interface with another material dramatically alters these states. This allows for the engineering of their electronic properties, such as their conductivity, by carefully selecting the adjacent material and the geometry of the junction~\cite{Rudi2024, Buccheri2024}. This control could be a pathway to novel electronic devices.
Additionally, the interactions between electrons within these engineered drumhead states can lead to the emergence of exotic collective phenomena.
For examples, placing an NLSM in proximity to a conventional superconductor is a particularly exciting direction. The drumhead states may inherit superconducting properties, potentially creating a topological superconductor. Such systems are predicted to host Majorana fermions, which are promising candidates for building fault-tolerant quantum computers. Recently, unconventional Andreev reflections in such systems have been explored in Refs~\cite{YXWang2022,XYWang2_2022}.

\item Quantum Geometry and Non-Linear Effects:
Quantum geometry describes the geometric properties of electron wavefunctions in momentum space, going beyond the simple band structure of energy versus momentum. It involves concepts like Berry curvature, which acts like a magnetic field in momentum space and fundamentally influences electron dynamics. One interesting effect is the non-linear Hall effect (NLHE). While the conventional Hall effect requires an external magnetic field, the NLHE can generate a transverse voltage purely from the material's intrinsic quantum geometry~\cite{Sodemann2015, Ma2019, Du2021, Lai2021, NWang2023,  Anyuan2023, Fujiwara2025, ulrich2025, HWang2025, Fang2024}.
The quantum geometric contribution to the NLHE can be significantly enhanced by (nearly gapped) Dirac/Weyl cones and nodal lines~\cite{Liao2024, LI2025, ulrich2025}. 
In NLSMs, the Berry curvature is expected to be highly concentrated around the nodal lines. Because these nodal lines form one-dimensional structures within the three-dimensional momentum space, the NLHE response is predicted to be strongly anisotropic, meaning it will depend heavily on the direction of the applied current relative to the orientation of the nodal lines. This makes NLSMs a fascinating platform for investigating and potentially harnessing quantum geometric effects for directional electronics.

\item Terahertz (THz) Applications:
The terahertz frequency ranges often called the "THz gap" because of the scarcity of efficient materials for generating, detecting, and manipulating radiation in this regime. NLSMs present two distinct advantages for overcoming this challenge.
One possible way to overcome this challenge is utilize the tunable narrow-gap platforms~\cite{Soranzio2024, Santhosh2024, Pettine2023, Rizza2022}. Optoelectronic devices like lasers and photodetectors require materials with an energy gap that matches the photon energy. For THz frequencies, this requires a very narrow gap. In an NLSM, the energy gap is zero along the nodal line but increases as you move away from it. By tuning the Fermi energy to lie very close to the nodal line, the material effectively behaves as a semiconductor with an inherently small and tunable gap, making it an ideal candidate for THz optoelectronics.
Another possibility source of non-linear THz generation is using the large concentration of Berry curvature near the nodal lines~\cite{Bera2023, Wang2022} . It can lead to strong non-linear interactions with light and  be exploited to generate THz radiation through processes like high-harmonic generation. While similar effects are studied in other topological materials like Weyl and Dirac semimetals~\cite{Kim2023, Rizza2022, Xi2025,  Ghalgaoui2025}, the unique line-node structure in NLSMs could offer different efficiencies and characteristics, making them a highly desirable platform for developing next-generation THz sources and detectors.
\end{itemize}

\begin{acknowledgments}
%P.-Y. C. 
The author
acknowledges support from
the National Science and Technology Council of Taiwan
under Grants No. NSTC 113-2112-M- 007-019, 114-2918-I-007-015, and the support from Yukawa Institute for Theoretical Physics,
Kyoto University,  RIKEN Center for Interdisciplinary Theoretical and Mathematical Sciences, and  National Center for Theoretical Sciences, Physics Division.
\end{acknowledgments}

\bibliography{reference}

\end{document}